\documentstyle[aps]{revtex}
\newcommand{\be}{\begin{equation}} \newcommand{\ee}{\end{equation}}
\newcommand{\bey}{\begin{eqnarray}} \newcommand{\eey}{\end{eqnarray}} 

\title{CP Violation in EPR-like neutrino oscillations} 
\author{M K Samal\footnote{e-mail:mksamal@iopb.ernet.in}} 
\address{Institute of Physics, Sachivalaya Marg, Bhubaneswar 751 005, India.} 

\begin{document} 
\maketitle 

\begin{abstract}
After reviewing the general framework to study EPR-like neutrino
oscillations, we derive expressions for the oscillation probabilities and
a direct measure of CP violation.  We compare the possibility of studying
CP violation in this case with that of baseline accelerator experiments
and conclude that it is possible to study CP violation in experiments with
length scales comparable to that of short baseline accelerator 
experiments.
\end{abstract} 

\vskip 3mm \centerline{\sf (Pre-print No: IP/BBSR/97-18, April 1997;
hep-ph/9707255.)}
\vskip 3mm 

Although the origin of CP violation is yet to be understood, the Standard Model
incorporates CP violation in the quark sector through the phase of CKM\cite{ckm}
mixing matrix. There is no such effect in its lepton sector because neutrinos are
massless by construct. However, at present the experiments involving solar neutrinos
and/or atmospheric neutrinos and the necessity of a hot component in the favoured
mixed Dark matter scenario indicate that neutrinos may possess a small mass. In such
a case, one would expect CP violation in massive neutrino oscillations, whose
experimental verification can help us understand the symmetry between the quark
sector and the lepton sector of the standard model.  Moreover, it is also
suggested\cite{fuku} that CP violation in lepton sector constitutes one of the key
ingredients of the mechanism for generating baryon asymmetry in the universe. 

The possibility of measuring CP violation in baseline accelerator neutrino
oscillations experiments has been analysed by various authors in recent
studies\cite{cpstud}. Though this CP violating effect is supressed in the
short baseline accelerator experiments if neutrinos have hierarchical mass
spectrum, it was pointed out that the suppression is avoidable in the long
baseline accelerator experiments, which is expected to operate in the near
future\cite{parkesuzuki}. However, the potential obstacle in measuring CP
violation in long-baseline neutrino experiments is the contamination due
to the matter effect. Since the earth matter is not CP symetric it can not
only produce fake CP violation which acts as contamination but also can
dominate over the CP phase effect in a certain region of the mixing
parameters in the $\nu_\mu \rightarrow \nu_e$ experiments\cite{conta}.

Hence it is reasonable to look for the possibility of studying CP violation in any
other experiments involving neutrino oscillations.  Recently it was
suggested\cite{dol} that in coincidence experiments, in which both the charged
lepton and neutrino, born in the same decay, are detected; specific EPR-like
neutrino oscillations\cite{einst} can show up. Beginning with the
suggestions\cite{lipkin} of using EPR-like effects to obtain information about
particle properties: the $\epsilon'$ parameter of $K^0$ decays and CP violation in
the $B$ meson systems, the study of EPR correlations has evolved from an
illustration of some of the surprising features of quantum mechanics to a practical
tool for determining physical parameters\cite{datta}.  In this paper
we study leptonic CP violation in such a scenario. 

In the following, after reviewing the general framework to study
EPR-like neutrino oscillations\cite{dol}, we derive expressions for the oscillation
probabilities and a direct measure of CP violation using the the modified Maiani
representation advocated by Particle Data Group\cite{pdg}.  Then we discuss the
possibility of studying CP violation in this case and compare it with that of Base
Line accelerator experiments. We conclude that it is possible to study CP violation
in experiments with length scales comparable to that of short baseline
accelerator experiments. 

If the neutrino mass matrix is not flavour diagonal then the flavour 
neutrinos $\nu_l \: (l=e, \mu, \tau)$ are non-trivial mixtures of the mass
eigenstates $\nu_j \: (j=1, 2, 3)$ with non-degenerate masses:
\begin{equation}
 \nu_l = \sum_j U_{lj} \nu_j.
\end{equation}
We consider the decay $\pi \rightarrow \mu \nu$ and the relevant part of the
standard model Lagrangian consists of the charged-current terms: 
\be
{\cal L}_\nu = g \sum_l \overline{l} \hat{W} \nu_l + H.c. = g \sum_{l,j} 
U_{lj} \overline{l} \hat{W} \nu_j + H.c.
\ee

We assume that all particles are on the mass shell:
\be
p_\pi=m_\pi^2, \: p_{\mu_j}^2=m_{\mu}^2, \: p_{\nu_j}=m_\nu^2,
\ee
and the pion has a definite 4-momentum $p_{\pi}= (E_\pi, {\bf p}_\pi)$
such that the conservation law:
\be
p_\pi= p_{\mu_j}+p_{\nu_j},  
\ee
alongwith the direction of, say, ${\bf p}_{\nu_j}$ determines the
4-momenta $p_{\mu_j}=(E_{\mu_j}, {\bf p}_{\mu_j})$ and
$p_{\nu_j}=(E_{\nu_j},{\bf p}_ {\nu_j})$. 

We analyse the type of experiments when both neutrino and muon are
detected. The probability to detect both muon $\mu$ and neutrino $\nu$ as
oscillating functions of the distance $d$ between the muon and neutrino
detection points $x_\mu$ and $x_\nu$ respectively, of the time interval
$\Delta t$ between ``clicks" of the two detector and the neutrino mixing
matrix elements is given as
\be
P_{\nu_l} (x_\mu, x_\nu)= | {\cal A}_{\nu_l} (x_\mu, x_\nu)|^2, 
\label{prob} 
\ee
where the amplitude to detect a muon at a space-time point $x_\mu$
together with a neutrino of flavour $l$ at a point $x_\nu$ is
\be
{\cal A}_{\nu_l}(x_\mu, x_\nu)= < \mu ; \nu_l | \psi_{p_\pi} (x_\mu, 
x_\nu)>, \label{amp}
\ee
where $\psi_{p_\pi}$ is the wavefunction of the $\mu \nu$ system.

For the two generations of Dirac neutrinos, the mixing matrix is devoid of 
any CP violating phase and the wavefunction of the $\mu \nu$ system evolves 
in space-time\cite{dol}   
\bey
\psi_{p_\pi}(x_\mu, x_\nu | x_i)
&=& |\mu> (|\nu_1> e^{-i \phi_1} \cos{\theta} + |\nu_2> e^{-i \phi_2} 
\sin{\theta}), 
\eey
where 
\bey
\phi_1 &=& p_{\nu_1} (x_\nu-x_i) + p_{\mu_1} (x_\mu-x_i),\nonumber\\
\phi_2 &=& p_{\nu_2} (x_\nu-x_i) + p_{\mu_2} (x_\mu-x_i),
\eey
and $|\mu>$, $|\nu_1>$ and $|\nu_2>$ are state vectors of the muon $\mu$ and 
neutrino mass eigenstates $\nu_1$ and $\nu_2$ respectively. The 
space-time co-ordinates of muon and neutrino are $x_\mu$ and $x_\nu$ 
respectively whereas $x_i$ is the co-ordinate of the decay point.
The muons emitted together with the $\nu_1$ and $\nu_2$ have respective 
momentum $p_{\mu_1}$ and $p_{\mu_2}$.
In this case the probability of detecting a $\nu_\mu$ is\cite{dol}
\be
P_{\nu_\mu}(x_\mu, x_\nu)= \cos^4{\theta} + \sin^4{\theta} + 2 
\sin^2{\theta} \cos^2{\theta} \cos(\phi_1-\phi_2). 
\ee
where 
\be
(\phi_1-\phi_2) = (p_{\nu_1}-p_{\nu_2})(x_\nu-x_\mu),
\ee
and the equality $(p_{\nu_1}-p_{\nu_2})=-(p_{\mu_1}-p_{\mu_2})$ has been 
used to get the above expression.
Similarly if the neutrino is detected to be $\nu_e$
\be
P_{\nu_e}(x_\mu, x_\nu)= 2 \sin^2{\theta} \cos^2{\theta} \{ 
1-\cos(\phi_1-\phi_2) \}. \label{2peu}
\ee
The probability oscillates in space and time with the change of $x_\mu$ 
and/or $x_\nu$, presenting a kind of EPR effect.
As we will see later, oscillation length and oscillation frequency are different from 
the standard values $L= 2 E_\nu / (m_1^2-m_2^2)$ and 
$L^{-1}$ respectively.

For the case of three generations of neutrinos, the probability (\ref{prob}) can be
written as
\bey
P_{\nu_\ell} (x_\mu, x_\nu)& =& 
\left| \sum_j U_{\ell j} U_{\mu j}^*
\mbox{e}^{-i\phi_j}\right|^2  \nonumber\\
& = & \sum_j |U_{\ell j}|^2 |U_{\mu j}|^2  
+ 2 \sum_{j<k} \mbox{Re} \, \left( U_{\ell j} U_{\mu j}^*
U_{\ell k}^* U_{\mu k} \right) \cos \phi_{jk}  + 2 \sum_{j<k}
\mbox{Im} \, \left( U_{\ell j} U_{\mu j}^* 
U_{\ell k}^* U_{\mu k} \right) \sin \phi_{jk}, \label{prob3}
\eey 
where $\phi_{jk}=(\phi_j-\phi_k)$.  The structure of the oscillation
formula is the same as that of standard neutrino oscillations except for
the nature of the phase. Before we get into the detailed discussion of the
phase $\phi_{jk}$ note that the coefficient of the $\sin \phi_{jk}$ in
eqn. (\ref{prob3}) is $2 J$, where $J$ is the rephasing invariant
Jarlskog's plaquette\cite{jarlskog} that is a measure of CP violation. To
isolate this term, it is easy to observe that
\be
P_{\nu_\ell} \left(x_\mu, x_\nu; \phi_{jk}= \frac{3
\pi}{2}\right)-P_{\nu_\ell} \left(x_\mu, x_\nu; 
\phi_{jk}= \frac{\pi}{2}\right)=4 J.  
\ee
Therefore measurement of $P_{\nu_\ell} \left(x_\mu, x_\nu; \phi_{jk}=
\frac{3 \pi}{2}\right)$ and $P_{\nu_\ell} \left(x_\mu, x_\nu; \phi_{jk}=
\frac{\pi}{2}\right)  $ can yield $J$.

The representation of the unitary mixing matrix depends
on the type of neutrinos. Majorana neutrinos can have three non-zero CP
phases whereas Dirac neutrinos can have only one CP phase. But it was
shown\cite{bilenky} that it is not possible to distinguish
experimentally between Dirac
and Majorana neutrinos using neutrino oscillations.  It is convenient to
describe the neutrino mixing matrix by the modified Maiani representation
advocated by Particle Data Group\cite{pdg}:
\be
\left(
\begin{array}{ccc} 
c_{12} c_{13}& s_{12} c_{13}&s_{13} e^{-i \delta}\\
-s_{12}c_{23}- c_{12} s_{13}s_{23} e^{i \delta}&c_{12}c_{23}- s_{12} 
s_{13}s_{23} e^{i \delta}  &c_{13} s_{23}\\ 
s_{12}s_{23}- c_{12} s_{13}c_{23} e^{i \delta} &-c_{12}s_{23}- s_{12} 
s_{13}c_{23} e^{i \delta}  &c_{13}c_{23} 
\end{array}
\right)
\ee
where $s_{jk}\equiv \sin{\theta_{jk}}$, $c_{jk}\equiv \cos{\theta_{jk}}$.  
In this parametrization, the expression for the probability of the neutrino 
being a $\nu_e$ is found out to be 
\be
P_{\nu_e}(x_\mu, x_\nu)= P_0 +P_1  + P_2 +P_3
,  \label{peu} 
\ee
where the coefficients $P_0$, $P_1$, $P_2$ and $P_3$ are functions of angles and
phases $\delta$ and $\phi_{jk}$ as: 
\bey
P_0&=& c_{13}^2 [ 2 c_{12}^2 c_{23}^2 s_{12}^2 +2 c_{12}^4 s_{13}^2 s_{23}^2 +(1+2
s_{12}^4) s_{13}^2 s_{23}^2 +2 c_{12} c_{23} s_{12} s_{23} s_{13}
(c_{12}^2-s_{12}^2) \nonumber\\
&& + 2 c_{12}^2 s_{12}^2 (s_{13}^2 s_{23}^2 - c_{23}^2)
\cos \phi_{12}], \nonumber\\
P_1&=& 2 c_{13}^2 c_{12} c_{23} s_{12} s_{23} s_{13} (\cos{\delta})
[(s_{12}^2-c_{12}^2) \cos \phi_{12} -\cos \phi_{13}+\cos \phi_{23}],
\nonumber\\
P_2&=& -c_{13}^2 s_{13}^2 s_{23}^2  (\cos{ 2 \delta}) [c_{12}^2 \cos \phi_{13}+
s_{12}^2
\cos \phi_{23}], \nonumber\\
P_3&= & 2 c_{13}^2 c_{12} c_{23} s_{12} s_{23} s_{13} (\sin{\delta}) [\sin \phi_{12}
-\sin \phi_{13}+\sin \phi_{23}] = J [\sin \phi_{12} -\sin \phi_{13}+\sin \phi_{23}].        
\eey

It is easy to check that the above expression reduces to the eqn.(\ref{2peu}) if one
puts $s_{23}=s_{13} \rightarrow 0$ and $c_{23}=c_{13} \rightarrow 1$ and $s_{12}
\equiv \sin{\theta}$. This probabilities oscillate not only with the
differences
(${\bf x}_\mu-{\bf x}_\nu) \equiv d $ and (${t}_\mu-{t}_\nu)\equiv \Delta t$ but
also with the CP phase $\delta$. The fluctuation of the envelope of this
probability with the KM phase $\delta$ can perhaps be observed in
experiments\cite{denisovsergi}.

To see the difference in the nature of phases, note that
for standard neutrino oscillations the phase $\phi_{jk}$ is
given as 
\be
\phi_{jk} = \frac{(m^2_{j}-m^2_k)}{2E} L=2.5 \: \left(\frac{\delta
m^2_{jk}}{10^{-2} \:
{\rm eV}^2}\right)\left(\frac{L}{100 \: km}\right)\left(\frac{E}{1 \:
{\rm GeV}}\right)^{-1},
\ee
where $L$ and $E$ are the source to detector distance and neutrino
energy respectively; whereas in the case of EPR-like neutrino oscillations\cite{dol}
\be
\phi_{jk} = \frac{\delta m^2_{jk}}{2E} t_{\nu i}, 
\ee
where $t_{\nu i}$ is the difference between the time of creation (i.e. the time of
pion decay) and the time of detection for the neutrino in natural units. The phase
does not depend on $t_{\mu i}$ and thus the above probability simply describes the
neutrino oscillations from the point of creation of the neutrino till the point of
its detection. Accordingly they do not depend on the position of the muon detector. 
Therefore EPR-like correlations between muon and neutrino detection appear only when
the time $t_{\nu i}$ is expressed in terms of $\Delta t=t_\nu -t_\mu$ and
$d=x_\mu-x_\nu$.  If one assumes that a special detector measures the decay point of
pion $x_i$ then the situation becomes absolutely trivial.

To express $t_{\nu i}$ in terms of $\Delta t$, 
consider a beam of pions moving from left to right along the line which connects the
muon and neutrino detectors. At fixed distance between the detectors $(d)$, the
measurement of $\Delta t$ (time difference between the clicks of the two detectors)
allow one to find the space-time point $x_i$ of the pion. Thus for a pion with
velocity $v_\pi > v^0_\mu$ (the muon velocity in pions rest frame) decaying to the
left of the muon detector we have\cite{dol} in the collinear case,
\be
t_{\nu i}= \frac{-v_\mu \Delta t + d}{v_\nu-v_\mu}. \label{gamma}
\ee 
On the otherhand a pion with velocity $v_\pi < v^0_\mu$, decaying between the two
detectors  
\be
t_{\nu i}= \frac{v_\mu \Delta t + d}{v_\nu+v_\mu}.
\ee

A direct measure of CP violation in three-neutrino oscillations is the unique
difference of the transition probabilities between CP-conjugate
channels\cite{sandip}:
\bey
\Delta P & \equiv & P(\overline{\nu}_\mu - \overline{\nu}_e) - 
P(\nu_\mu - \nu_e) = P(\nu_\mu - \nu_\tau) - P(\overline{\nu}_\mu - 
\overline{\nu}_\tau) = P(\overline{\nu}_e - \overline{\nu}_\tau)  - 
P(\nu_e - \nu_\tau) \nonumber\\
&= & 4 J S,
\eey 
where the contribution from the mixing matrix is in terms of $J$ and the
squared mass differences contribute through the term $S=\sum_{j < k} \sin
\phi_{jk}$. 
To estimate $\Delta P$ one uses the clues about the $J$ and $S$ from the
experiments regarding neutrino masses and mixing.

There are only two hierarchical mass difference scales $\delta m^2$ in the
three-flavour mixing scheme. If the highest neutrino mass scale is taken
to be $O(1\: {\rm eV})$, which is appropriate for cosmological hot dark
matter\cite{cardell}, the other mass scale is either the atmospheric
neutrino mass scale $\delta m^2 \simeq 10^{-2} \:\;{\rm eV}^2$ or the
solar neutrino mass scale $\delta m^2 \simeq 10^{-5} \sim 10^{-6}\:\; {\rm
eV}^2$.  The lower mass scale is chosen to be $\delta m^2 \simeq 10^{-2}
\:\; {\rm eV}^2$ for studies of CP violation in long baseline experiments
as they correspond to the atmospheric neutrino mass scale. Then, one
should introduce the sterile neutrino to solve the solar neutrino puzzle
via neutrino oscillations. However, all the neutrino data can be
consistent with an almost degenerate neutrinos scenario and CP violation
in this case has been studied\cite{arf}. 
Once the neutrino mass scales are fixed, one
can discuss the pattern of the $3 \times 3$ neutrino mixing matrix. In
general, there are three allowed regions of the mixings, which are derived
from the reactor and accelerator disappearance experiments and $J$ is
estimated\cite{tani} accordingly. 

In the case of EPR-like oscillations, one gets the same $J$ contribution
as that of standard oscillations, but a different contribution for $S$. 
In standard oscillations, one obtains for short baseline experiments
$\phi_{31} \simeq -\phi_{12} = O(1)$ if $\Delta m^2_{31} = O(1 \: {\rm
eV}^2)$. The factor $S$ is supressed because the two largest terms almost
cancel eachother due to their opposite signs and the term containing
$\phi_{23}$ is small. Due to this significant reduction of the term $S$ it
is argued that one has no chance to observe CP violation in the short
baseline neutrino oscillation experiments. However, the situation is
different in the long baseline accelerator experiments. Since the
magnitude of $\phi_{31} (\simeq -\phi_{12}) $ is $10^3 \sim 10^4$, the
terms containing them average out to be zero, but at the same time the
term containing $\phi_{23}$ is $O(1)$. Thus the quantity $\Delta P$ is not
supressed unless $J$ is too small.

To discuss the contributions to $S$ in the case of EPR-like oscillations,
let us note that the energy scale involved is of tens of MeV. Hence for
the neutrino mass hierarchy chosen above, $S$ will get a substantial
contribution through $\phi_{23}$ if $t_{\nu i}$ can be hundreds of meters.
For the case $v_\pi < v_\mu^0$ this can be achieved by chosing the
separation between the detectors ($d$) to be few hundreds of meters, which
is the same scale as short baseline experiments.  But in the case of pion
with $v_\pi >v_\mu^0$, this can be achieved with $d$ of tens of meters
because for the ultra-relativistic muons with the denominator in the
expression (\ref{gamma}) could be very small and hence $t_{\nu i}$ can be
enhanced. But if the numerator in this case is also very small then the
contribution to $S$ gets suppressed. 

To conclude, we have shown that unlike the case of base line neutrino
oscillation experiments, it is possible to study CP violation in
EPR-like neutrino oscillations experiments with length scales
comparable to that of short base line experiments. Hence in this case one
does not have to bother about the contamination due to matter effect.
We also give the expressions for the oscillation probabilities whose
envelopes' fluctuation with the KM phase $\delta$ can tell us about the
magnitude of CP violation.

\section*{Acknowledgements}
I am grateful to Subhendra Mohanty for drawing my attention to ref. 
\cite{dol} and for discussions. I am thankful to Theory Group at 
Physical Research Laboratory, Ahmedabad for local hospitality during a visit 
when this work was initiated. It is a pleasure to thank the referee for
useful comments.

\end{document}